# WIRE SCANNERS FOR SMALL EMITTANCE BEAM MEASUREMENT IN ATF

H. Hayano, KEK, Tsukuba, Ibaraki, Japan

*Abstract*

The wire scanners are used for a measurement of the very small beam size and the emittance in Accelerator Test Facility (ATF). They are installed in the extraction beam line of ATF damping ring. The extracted beam emittance are $\varepsilon_x$=1.3x$10^{-9}$ m.rad, $\varepsilon_y$=1.7x$10^{-11}$ m.rad with 2x$10^{9}$ electrons/bunch intensity and 1.3GeV energy[1]. The wire scanners scan the beam by a tungsten wire with beam repetition 0.78Hz. The scanning speed is, however, very slow( ~500µm/sec). Since the extracted beam is quite stable by using the double kicker system[2], precision of the size measurement is less than 2µm for 50 - 150µm horizontal beam size and 0.3µm for 8 - 16µm vertical beam size. The detail of the system and the performance are described.

## 1 INTRODUCTION

ATF is a test accelerator to realize a small emittance beam which will be used in an electron positron linear collider. The beam emittance measurement in the ATF extraction line is required for a single bunch and 20 multi-bunched beam which has 2x$10^{10}$ electrons in each bunch with 2.8ns spacing[3]. The required resolution of the beam size monitors is less than 1µm for the beam of 6 - 7µm vertical size. On the other hand, the horizontal beam size is around 50 -150µm, bigger than vertical one. The wire scanner beam size monitor is the most appropriate monitor for the required performance. The five wire scanners together with a gamma detector at downstream are installed at the region of no dispersion in the extraction line between qudrupole magnets. The beam size from 4 or 5 scanners are used to fit an emittance assuming the optics between the scanners[4].

In order to measure precisely such a small vertical beam size with big horizontal size, the installation of the wire in the beam line and precision of the movement are very important. The pulse-to-pulse stability of the beam is also important for such a scanning monitor. The performance of the wire installation and movement are described together with measurement stability.

## 2 WIRE SCANNER

Fig.1 shows the picture of the wire mover stage and the vacuum chamber. The wire mount shown in Fig.2 is supported by the two arms in the vacuum chamber. A 50µm diameter gold plated tungsten wire is stretched simultaneously to X, Y and U directions which is 45 degree tilted from the X direction. In the other side of the mount, a 10µm diameter gold plated tungsten wire is also stretched to X, Y, U and 10degree tilted from X direction. The only one wire is directly stretched on the steel mount between precise steel pillars in order to have a precision of the stretched angle. Both end of wire are fixed to copper pillars with holding by solder. The moving direction of the wire stage is 45 degree between X and Y axis has an advantage of that the only one move direction is necessary for 3 axis scans. The angles of the stretching wires are measured by using a microscope with precision mover stage. An example of the measurement is shown in Fig.3. Each measured point are fitted to a linear function and relative angles are calculated. The results are summarized in Table 1. The error from the design angle is less than 1 degree. The installation into the beamline was done to have 10μm horizontal wire to sit in the precise horizontal. Using an alignment telescope, tilt angle of the whole wire scanner chamber was adjusted within 0.2 degree.

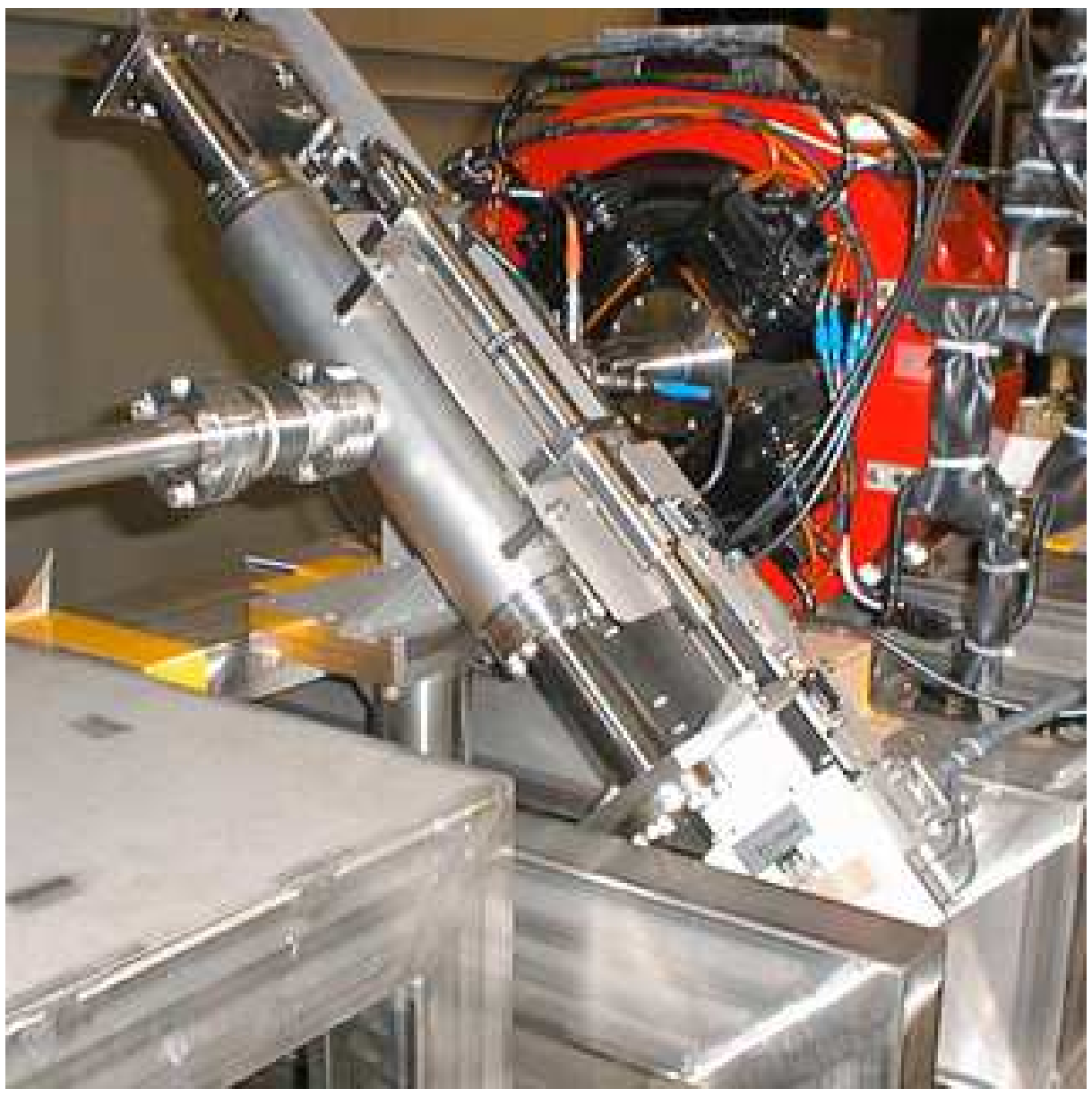

Fig.1: Wire scanner chamber in ATF extraction line.

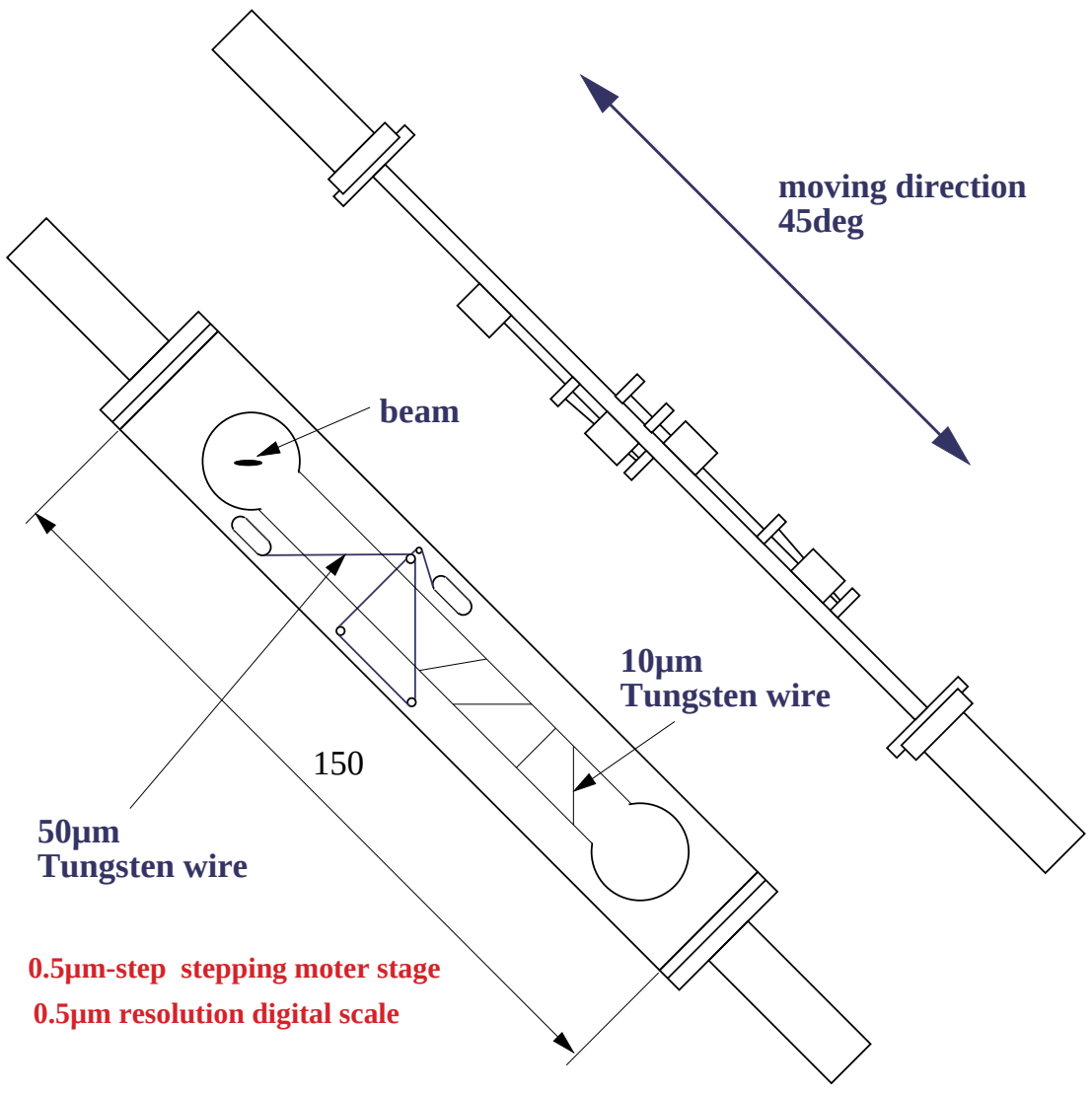

Fig.2 Wire mount in the wire scanner chamber

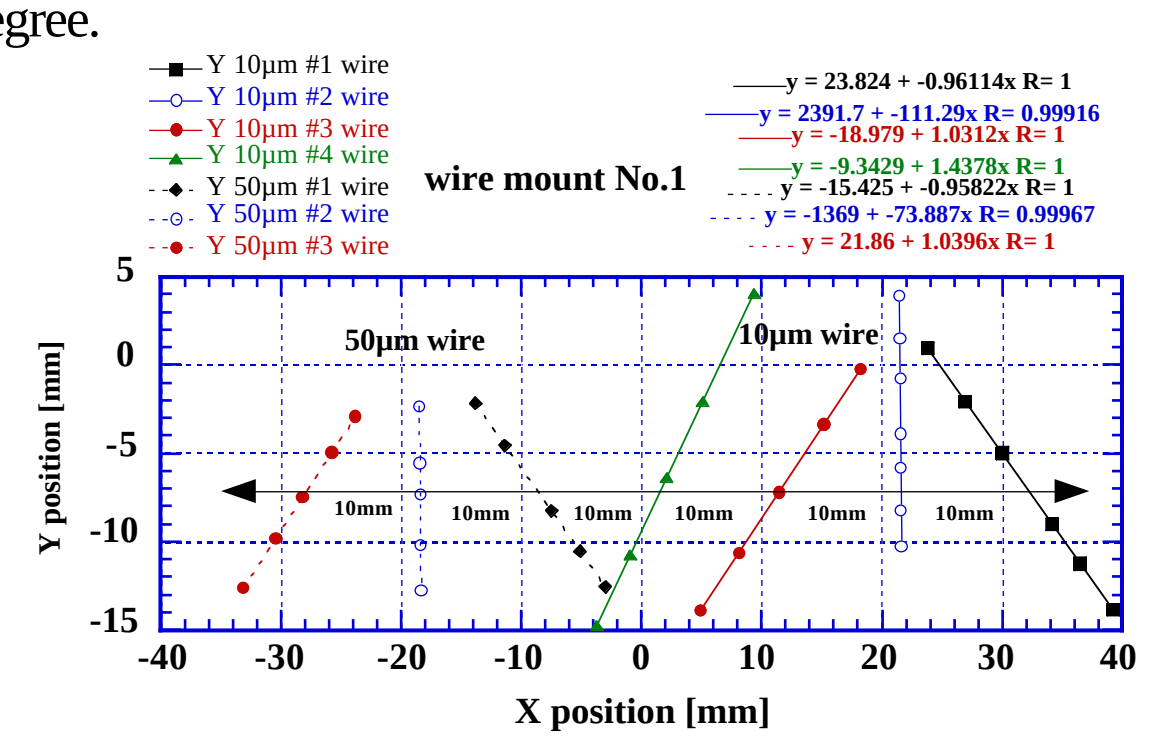

Fig.3:measurement of wire stretching angle.

Table 1: measured stretch angle of wire

| stretch angle [degree] | No.1 wire mount | No.2 wire mount | No.3 wire mount | No.4 wire mount |
|---|---|---|---|---|
| 10μm#1 wire | -89.745 | -90.614 | -90.074 | -90.062 |
| #2 | -135.365 | -134.849 | -134.994 | -134.942 |
| #3 | 0.000 | 0.000 | 0.000 | 0.000 |
| #4 | 9.301 | 9.276 | 9.628 | 9.705 |
| 50μm#1 wire | -89.658 | -90.534 | -90.322 | -90.211 |
| #2 | -135.104 | -135.626 | -134.845 | -134.978 |
| #3 | 0.232 | -0.694 | -0.099 | -0.069 |

The two arm supports of the wire mount inside the vacuum chamber are fixed to the two stages with two bellows at both ends of support arm tubes. The support tubes are loaded by the vacuum pressure loading from both side of tubes, however, the pressure cancels it out on the movement of the stage. The load of mover stage is a spring recovery force of the bellows only. Furthermore a vibration of the wire mount is reduced by using this double support stage compared with a single end support. As it has many advantages, the double stage mover was adopted. The one end of the wire mover stage is powered by a 5-phase stepping motor stage assembly (Physik Instrumente M-510.10). This stage is driven by a ball bearing spindle of 1mm pitch for one revolution. The stepping motor performs one rotation by 2000 steps. Combining the ball bearing spindle and the precise stepping motor, the resolution of step is 0.5μm and repeatability is less than 0.1μm. As a maximum starting pulse rate is limited to 1 kHz, a 1024 pulse/sec constant pulse rate is used for the stage control for the reason of simplicity.

The stepping motor stage assembly has a radiation resistant. Also, the stage position sensor must have a radiation resistant. The Magnescale position sensor is adopted, because it does not have processing electronics near the sensor. It include a magnetized rod with very fine pitch and a pickup coil. The processing electronics is placed outside of the accelerator tunnel. The resolution of the Magnescale is 0.5μm for 100mm travel, enough small for the beam size measurement of 10μm.

The vibration measurements[5] were done for the same type of the stage using a laser light beam on the wire. The cw laser beam of 70μm diameter was used to simulate stable particle beam. The vibration of wire appears on the absorption change of the photo-detector at downstream of the laser beam. The observed vibration of wire( 50μm diameter ) was always less than 0.3μmp.p. for 55 Hz to 771 Hz clock speed. With higher clock speed for the stepping motor more than 150 Hz reduce the vibration amplitude to 0.2μmp.p.. It is enough small amplitude for ATF beam measurement.

## 3 WIRE SIGNAL DETECTION

A breamsstrahlung gamma-ray is used for the signal of beam and wire interaction. A gamma-ray detector[6] is placed at the last bending magnet in the end of beam diagnostic section. The detector is a cerencov detector which consists of 2mm thick lead plate converter, air light guide and a photo-multiplier tube(PMT). In order to shield the PMT from other noise source such as gamma-ray and neutrons caused by beam loss in other place, the PMT is placed at the floor with lead shield and the 1m long light guide is used between the cerenkov light radiator and the PMT. Since the gamma detection is done by a calorimetric way, low HV voltage is applied to the PMT with light intensity filter in front of the PMT. The intensity filter is made by thin plate with many small holes which reduce the light intensity. The HV voltage is changed from 600V to 950V depending on the beam size and the wire diameter. The wire signal out from the PMT is a pulse signal with around 20ns width detected by a charge sensitive ADC using 100ns gate. Since the ADC is installed at the outside of the tunnel, a 20m coaxial cable is used between the PMT and the ADC.

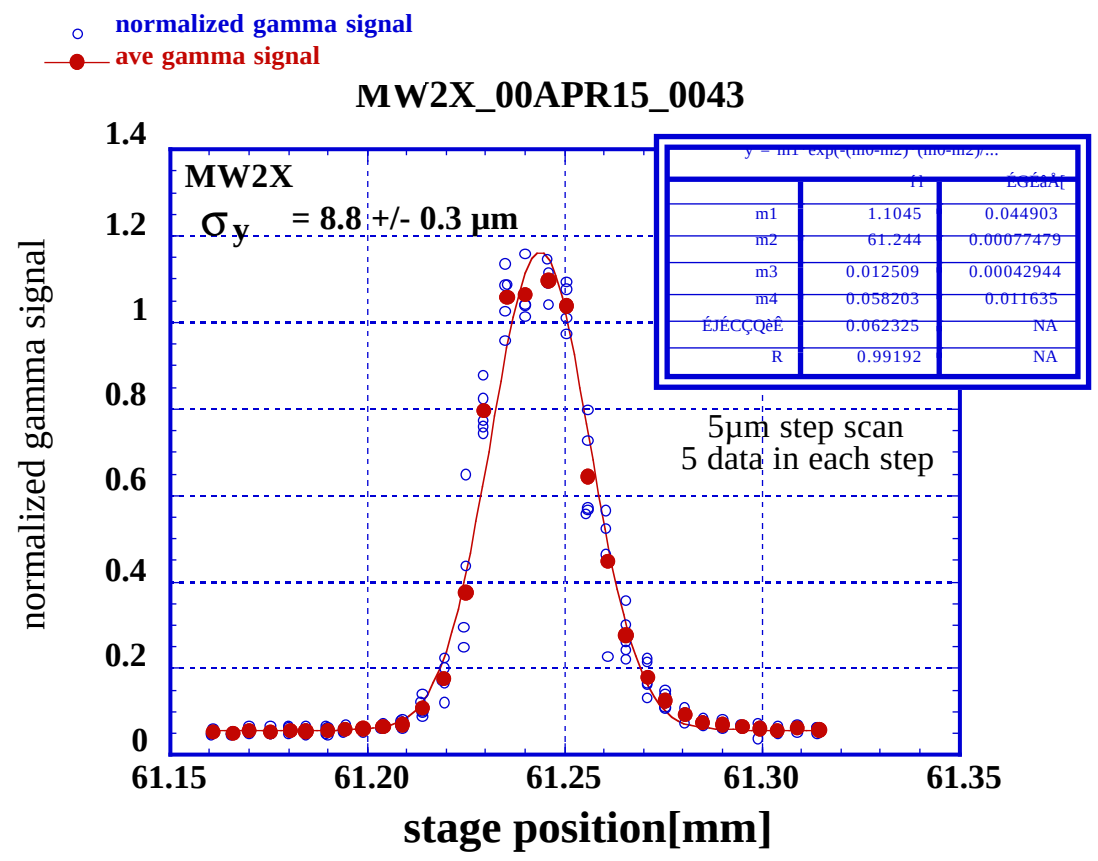

Fig.4: Example of scanned beam profile.

## 4 BEAM SIZE MEASUREMENT

The acquisition of the wire signal is done by the BPM reader task which read the all BPM signal and the beam intensity monitor synchronously with one ATF beam cycle. As a consequence, the wire signal, the beam intensities and the beam positions are stored into the database for the same beam pass. A fluctuation of the wire signal caused by the beam intensity is corrected by normalizing the wire signal by the extraction beam intensity. As an example, a fluctuation of wire signal is shown in Fig.4 together with its averaged signal. The spread of the distribution which is not so much gives 0.3µm error in the fitted beam size. A position fluctuation is not corrected, because of insufficient resolution of BPM compared with beam fluctuation. The resolution of BPM is around 10 to 20µm, while the position jitter of beam is estimated to 2.4 to 4.4µm by using the wire sitting in beam and using high resolution microwave BPMs which are newly installed in the same region.

In case of X, Y and 10 degree wires, stage positions are recorded as an abscissa which must be converted to a real movement in its directions, such as multiplying 1/sqrt(2) for X and Y. The size also must be corrected by the wire diameter. The effect is quadratic of (d/4) where d is a wire diameter. The simulation shows that the corrected beam size is enough accurate down to 2µm beam size.

The scanning speed is around 30sec for one profile. The slow scan is coming from both the 0.78Hz beam repetition and the slow stage movement. The scan is made as follows; by setting the wire stage to the initial position at first, then wait the beam passing, get the wire signal together with the beam intensity, go to the next stage position, then repeat again until the end position coming. This means that the shortest scanning time is the beam repetition times number of the scanning points.

## 5 EMITTANCE MEASUREMENT

Beam size measurements at more than three locations in different optical condition are necessary for measuring an emittance. The optics of the wire scanner region is determined by the SAD simulation to have less error in the emittance calculation assuming reasonable measurement error[7]. In case of ATF flat beam, vertical beam size measurement is affected by a small beam tilt or a small wire tilt. The beam tilt happens by an x-y coupling and a residual x, y dispersion. Four skew quadrupole magnets are introduced into the upstream of the wire scanner region to compensate beam tilt. Also, the careful tuning of dispersion suppression less than 10mm is necessary for the vertical emittance measurement. A 10mm dispersion causes about 6µm beam size increment which is comparable to the vertical size. The dispersion is measured by wire scanner with 1mm resolution using peak shift of the scanned profile by changing the ring frequency. The measured Y emittance by changing the beam intensity using one skew quadrupole are shown in Fig.5. The observed emittance growth with beam intensity is larger than the growth of an intrabeam effect. The further study on this emittance growth is now in progress.

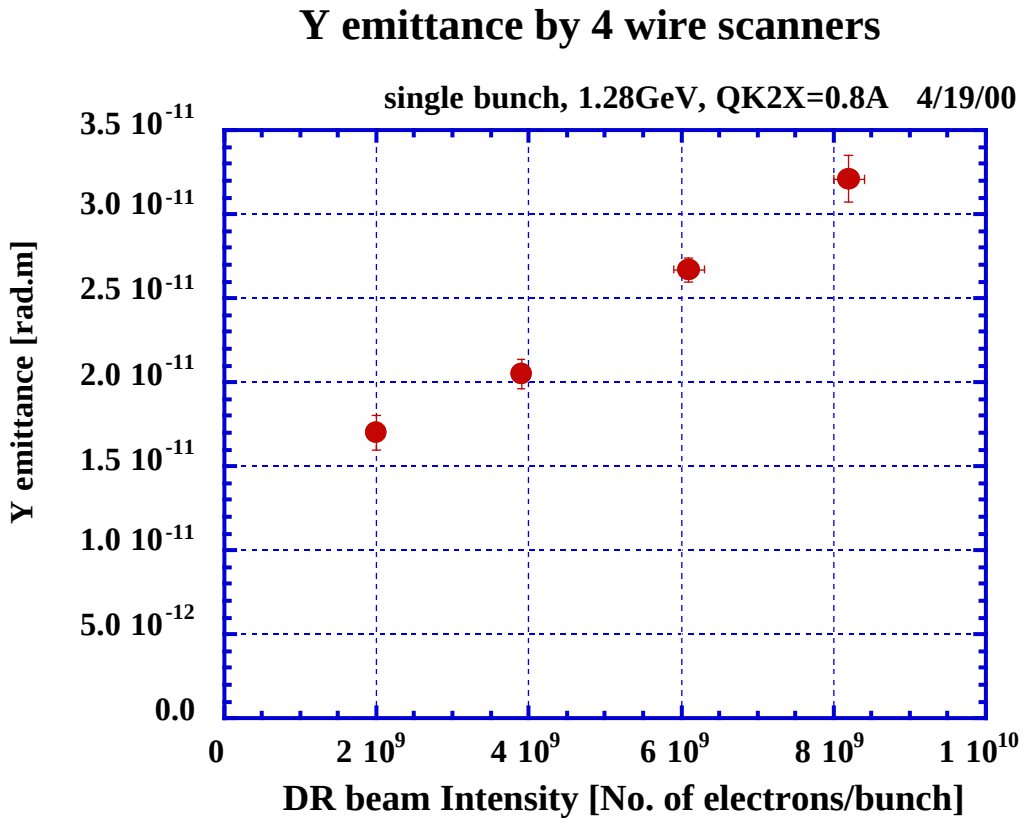

Fig.5: measured Y emittance vs. beam intensity

## 6 SUMMARY

The five wire scanners in the ATF extraction line are used to measure the beam size of 50 - 150µm in horizontal and 8 - 16µm in vertical. Since the extracted beam is stable within 4.4µm, the wire scanner measurement is performed by less than 2µm stability in X and 0.3µm in Y. The measured emittance are near the ATF target in case of low beam intensity. The emittance growth with beam intensity is still not understandable.

## 7 ACKNOWLEDGMENT

The author would like to acknowledge Prof. Sugawara, director of KEK organization, Prof. Kihara, director of Accelerator research, Prof. Iwata, Prof. Takata and Prof. Yamazaki for their support of ATF. The author also thank to M. Ross, D. McCormick of SLAC and the member of ATF for their cooperation and useful discussion.